\documentclass[letterpaper]{article} % DO NOT CHANGE THIS
\usepackage{aaai2026}  % DO NOT CHANGE THIS
\usepackage{times}  % DO NOT CHANGE THIS
\usepackage{helvet}  % DO NOT CHANGE THIS
\usepackage{courier}  % DO NOT CHANGE THIS
\usepackage[hyphens]{url}  % DO NOT CHANGE THIS
\usepackage{graphicx} % DO NOT CHANGE THIS
\usepackage{natbib}  % DO NOT CHANGE THIS AND DO NOT ADD ANY OPTIONS TO IT
\usepackage{caption} % DO NOT CHANGE THIS AND DO NOT ADD ANY OPTIONS TO IT
\usepackage{algorithm}
\usepackage{algorithmic}
\usepackage{amsmath}
\usepackage{amsthm}
\usepackage{tabularx}
\usepackage{multirow}
\usepackage{xcolor} % thêm ở phần preamble

\usepackage{newfloat}
\usepackage{listings}
\DeclareCaptionStyle{ruled}{labelfont=normalfont,labelsep=colon,strut=off} % DO NOT CHANGE THIS
\floatstyle{ruled}
\newfloat{listing}{tb}{lst}{}
\floatname{listing}{Listing}
\title{Graph-Theoretic Consistency for Robust and Topology-Aware Semi-Supervised Histopathology Segmentation (Student Abstract)}
\author{
    Ha-Hieu Pham\textsuperscript{\rm 1,2,3},
    Minh Le\textsuperscript{\rm 4},
    Han Huynh\textsuperscript{\rm 4},
    Nguyen Quoc Khanh Le\textsuperscript{\rm 4}, 
    Huy-Hieu Pham\textsuperscript{\rm 3,5}\textsuperscript{*}% corresponding 
}
\affiliations{
    \textsuperscript{\rm 1}University of Science, VNU-HCM, Ho Chi Minh City, Vietnam\\
    \textsuperscript{\rm 2}Vietnam National University, Ho Chi Minh City, Vietnam\\
    \textsuperscript{\rm 3}VinUni-Illinois Smart Health Center, VinUniversity, Hanoi City, Vietnam\\
    \textsuperscript{\rm 4}AIBioMed Research Group, Taipei Medical University, Taipei City, Taiwan\\
    \textsuperscript{\rm 5}College of Engineering \& Computer Science, VinUniversity, Hanoi City, Vietnam\\
    hieu.ph2@vinuni.edu.vn, \{D142111009, m658112001, Khanhlee\}@tmu.edu.tw, hieu.ph@vinuni.edu.vn \\
    {*}Corresponding author: hieu.ph@vinuni.edu.vn
}

\begin{document}

\maketitle

\begin{abstract}
Semi-supervised semantic segmentation (SSSS) is vital in computational pathology, where dense annotations are costly and limited. Existing methods often rely on pixel-level consistency, which propagates noisy pseudo-labels and produces fragmented or topologically invalid masks. We propose \textbf{Topology Graph Consistency (TGC)}, a framework that integrates graph-theoretic constraints by aligning Laplacian spectra, component counts, and adjacency statistics between prediction graphs and references. This enforces global topology and improves segmentation accuracy. Experiments on GlaS and CRAG demonstrate that TGC achieves state-of-the-art performance under 5--10\% supervision and significantly narrows the gap to full supervision.
\end{abstract}

% \section{Introduction}

% Semantic segmentation in histopathology is critical for computational pathology, but dense pixel-level annotations are costly and time-consuming. Semi-supervised segmentation helps reduce annotation effort by using unlabeled data, yet most existing methods rely on pixel-level consistency, which amplifies noise and produces fragmented or topologically invalid masks, especially problematic when gland morphology is clinically relevant \cite{pham2025fetal, csds, vu2025semi, hdc, nguyen2024blurry}. Unlike pixels, graphs naturally model region relationships and global structure, making them well-suited for preserving biological topology such as gland connectivity or lumen enclosure \cite{graph}. Motivated by this, we propose \textbf{Topology Graph Consistency (TGC)}, a dual-network framework that converts predictions into region-level graphs and aligns them with references through spectral and structural constraints. By enforcing global topological alignment in addition to local accuracy, TGC improves segmentation performance and preserves meaningful structure. Experiments on GlaS and CRAG ~\cite{SIRINUKUNWATTANA2017489, GRAHAM2019199} show that TGC achieves state-of-the-art results under 5--10\% supervision.

% Add this just before copyright
\section{Introduction}

Semantic segmentation in histopathology is a key step for computational pathology, enabling analysis of tissue architecture and cancer grading. Yet, obtaining dense pixel-level annotations remains costly and requires expert effort, especially for complex glandular structures in colorectal tissues. Semi-supervised learning provides a practical solution by leveraging unlabeled data to improve segmentation with limited supervision. However, most existing methods enforce pixel-level consistency, which is prone to label noise and tends to produce fragmented or topologically invalid masks. These errors are particularly critical when gland morphology and lumen enclosure have diagnostic importance.

Unlike pixels, region-level representations can model structural relationships between tissue components. Graph-based formulations are thus well suited for histopathology, as they naturally capture connectivity, adjacency, and topology between glandular regions \cite{graph, kipf2017gnn, zhang2020gnnvision}. By reasoning over region graphs instead of individual pixels, segmentation models can preserve biologically meaningful topology and reduce the risk of inconsistent gland boundaries.

Motivated by this, we propose \textbf{Topology Graph Consistency (TGC)}, a dual-network semi-supervised framework that converts segmentation maps into region-level graphs and aligns them through spectral and structural constraints. TGC encourages both local accuracy and global topological coherence, resulting in more robust and morphologically faithful predictions. Experiments on the GlaS \cite{SIRINUKUNWATTANA2017489} and CRAG \cite{GRAHAM2019199} datasets show that TGC achieves state-of-the-art results under 5--10\% supervision, outperforming recent semi-supervised methods while better preserving gland topology.

\begin{figure}[t]
    \centering
    \includegraphics[width=1\linewidth]{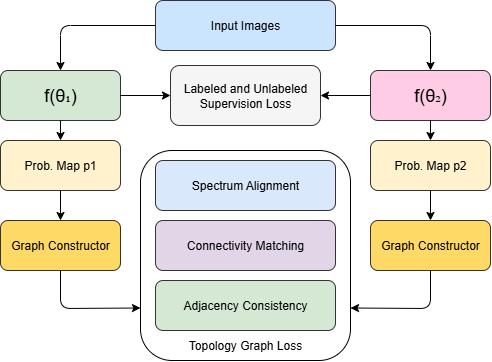}
    \caption{Overview of the proposed TGC framework. Two networks $f(\theta_1), f(\theta_2)$ process labeled and unlabeled inputs, producing probability maps converted into graphs. Graph descriptors (spectrum, connectivity, adjacency) define the topology loss, complementing DiceCE supervision on labeled data and pseudo-label consistency on unlabeled data.}
    \label{fig:placeholder}
\end{figure}

\section{Methodology}

An overview of the proposed \textbf{Topology Graph Consistency (TGC)} framework is illustrated in Figure~\ref{fig:placeholder}. 
We design a dual-network semi-supervised segmentation strategy with a novel topology-aware loss, 
which augments pixel-level supervision with graph-theoretic constraints to enforce structural plausibility.

\vspace{0.3em}
\noindent\textbf{Labeled and Unlabeled Supervision.}  
Two models $f(\theta_1), f(\theta_2)$ of the same architecture are trained jointly with Dice+CE loss.  
For an input $x$ with reference label $y$ (ground truth if labeled, or pseudo-label from the other model if unlabeled),  
the loss is:  
\begin{equation}
\begin{aligned}
\mathcal{L}_{DiceCE}(x,y) 
&= \mathcal{L}_{DiceCE}(f(\theta_1; x), y) \\
&\quad + \mathcal{L}_{DiceCE}(f(\theta_2; x), y).
\end{aligned}
\end{equation}
This unified formulation covers both labeled $(x_l,y_l)$ and unlabeled $(x_u,\hat{y})$ cases,  
where $\hat{y}$ denotes the pseudo-label exchanged between models, reducing confirmation bias.

\vspace{0.3em}
\noindent\textbf{Graph Construction.}  
Given a probability map $p \in [0,1]^{H\times W}$, we extract centroids $\{c_i\}$ to represent gland regions and build a $k$-nearest neighbor graph. The adjacency, degree, and Laplacian are:
\begin{equation}
\begin{aligned}
A_{ij} &=
\begin{cases}
\exp\!\left(-\tfrac{\|c_i-c_j\|^2}{2\sigma^2}\right), & j \in kNN(i), \\
0, & \text{otherwise},
\end{cases} \\
D &= \mathrm{diag}(A\mathbf{1}), \quad 
L = I - D^{-\tfrac{1}{2}} A D^{-\tfrac{1}{2}}.
\end{aligned}
\end{equation}
Here $c_i, c_j$ are centroid coordinates, $\sigma$ controls affinity decay, $k$ is the neighborhood size, $A$ is the weighted adjacency, $D$ the degree matrix, and $L$ the normalized Laplacian.

\vspace{0.3em}
\noindent\textbf{Topology Graph Loss.}  
Given prediction graph $G_p$ and reference graph $G_r$, we define:
\begin{equation}
\mathcal{L}_{spec} = \tfrac{1}{m-1}\sum_{i=2}^m (\lambda_i^{(p)}-\lambda_i^{(r)})^2,
\end{equation}
where $\lambda_i$ are Laplacian eigenvalues.  
\begin{equation}
\mathcal{L}_{conn} = (\hat{k}(G_p)-\hat{k}(G_r))^2, \quad
\hat{k}(G) = \sum_{i=1}^m \sigma((\tau-\lambda_i)\alpha),
\end{equation}
where $\tau$ is a threshold and $\alpha$ a sharpness factor.  
\begin{equation}
\mathcal{L}_{adj} = 
\tfrac{1}{\min(N_p,N_r)} \|\text{sort}(D_p)-\text{sort}(D_r)\|_2^2 +
(\bar{A}_p-\bar{A}_r)^2.
\end{equation}
Here $D_p, D_r$ are degree vectors and $\bar{A}$ the mean adjacency.  

The total topology loss is:
\begin{equation}
\mathcal{L}_{TGC} = w_{spec}\mathcal{L}_{spec} + w_{conn}\mathcal{L}_{conn} + w_{adj}\mathcal{L}_{adj},
\end{equation}
with $w_{spec}, w_{conn}, w_{adj}$ as balancing weights.

\vspace{0.3em}
\noindent\textbf{Total Objective.}  
The complete objective integrates pixel and graph-level supervision:
\begin{equation}
\mathcal{L}_{total} =
\mathcal{L}^{(l)}_{DiceCE} + \lambda_{sup}\mathcal{L}^{(l)}_{TGC} +
\mathcal{L}^{(u)}_{DiceCE} + \lambda_{unsup}\mathcal{L}^{(u)}_{TGC}.
\end{equation}
Here $\lambda_{sup}, \lambda_{unsup}$ control the strength of topology regularization, with $\lambda_{unsup}$ ramped up during training to mitigate early pseudo-label noise.

\section{Experiments}

\begin{table}[t]
\centering
\begin{tabular}{l l l cc}
\hline
\textbf{Dataset} & \textbf{Ratio} & \textbf{Method} & \textbf{Dice} & \textbf{Jaccard} \\
\hline
\multirow{8}{*}{GlaS} 
 & \multirow{4}{*}{5\%}
   & CCVC & 80.8 & 68.9 \\
 &  & CorrMatch & 79.9 & 67.8 \\
 &  & FDCL & \underline{81.6} & \underline{70.2} \\
 &  & \textbf{Ours} & \textbf{82.7} & \textbf{71.8} \\
\cline{2-5}
 & \multirow{4}{*}{10\%}
   & CCVC & 83.8 & 73.5 \\
 &  & CorrMatch & 83.3 & 72.6 \\
 &  & FDCL & \underline{84.4} & \underline{74.5} \\
 &  & \textbf{Ours} & \textbf{85.2} & \textbf{74.8} \\
\hline
\multirow{8}{*}{CRAG}
 & \multirow{4}{*}{5\%}
   & CCVC & 73.3 & 60.5 \\
 &  & CorrMatch & 69.1 & 55.4 \\
 &  & FDCL & \underline{74.6} & \underline{61.9} \\
 &  & \textbf{Ours} & \textbf{75.1} & \textbf{62.9} \\
\cline{2-5}
 & \multirow{4}{*}{10\%}
   & CCVC & 75.0 & 62.3 \\
 &  & CorrMatch & 74.9 & 61.9 \\
 &  & FDCL & \underline{76.3} & \underline{63.9} \\
 &  & \textbf{Ours} & \textbf{79.6} & \textbf{67.9} \\
\hline
\end{tabular}
\caption{Results on GlaS and CRAG (Dice/Jaccard, \%). Values are the mean over 5-fold cross-validation. Best in \textbf{bold}, second-best underlined.}
\label{tab:glas_crag_results}
\end{table}

\noindent\textbf{Implementation.} 
We implement all experiments in PyTorch on a single NVIDIA RTX 3060 GPU (16GB), using DeepLabV3+ with a ResNet-101 backbone. Models are trained for 80 epochs on GlaS and 120 epochs on CRAG, with all images resized to $256 \times 256$. Standard data augmentations (random flips and rotations) are applied. We use AdamW optimizer (learning rate $1 \times 10^{-4}$, weight decay $0.05$), batch size 8, and select the best validation checkpoint for inference. Our Topology Graph Consistency (TGC) loss is integrated into the dual-network training objective. All results are averaged over 5-fold cross-validation.

\noindent\textbf{Results.} 
As shown in Table~\ref{tab:glas_crag_results}, the proposed TGC framework consistently achieves top performance under both 5\% and 10\% supervision, outperforming or matching recent methods such as CCVC~\cite{ccvc}, CorrMatch~\cite{corrmatch}, and FDCL~\cite{fdcl}. In addition to quantitative improvements, qualitative results show that TGC yields fewer fragmented predictions and better preserves glandular structures.

\section{Conclusion and Future Work}

We proposed \textbf{Topology Graph Consistency (TGC)}, a semi-supervised segmentation framework that enforces topological alignment through graph-based constraints. By incorporating spectral, connectivity, and adjacency cues, TGC encourages structurally consistent predictions even under limited supervision. This approach highlights the importance of integrating topological priors into segmentation models, particularly in medical imaging, where structural correctness is essential.

In future work, we aim to expand TGC to support various imaging modalities and anatomical structures. We also plan to explore more expressive graph formulations, such as hypergraphs, to model higher-order spatial relations. Moreover, we will investigate the use of topology-aware GNNs and attention-based modules to further improve segmentation performance and robustness.

\section{Acknowledgment}
This research was supported by research funding from the VinUni-Illinois Smart Health Center, VinUniversity, Ha Noi City, Vietnam.

\bibliography{aaai2026}

\end{document}